\documentstyle[11pt,subeqn]{article}
\renewcommand{\baselinestretch}{1.5}

\begin{document}
\begin{titlepage}
\large
\begin{center}
{\bf Gravitational Waves from an Axi--symmetric Source in the
Nonsymmetric Gravitational Theory\\}
\vspace{.4in}
{\bf J. W. Moffat and D. C. Tatarski\\}
{\bf Department of Physics\\}
{\bf University of Toronto\\}
{\bf Toronto, Ontario M5S 1A7, Canada\\}
\vspace{0.4in}
{\bf Abstract}
\end{center}
\normalsize

We examine gravitational waves in an isolated axi--symmetric
reflexion symmetric NGT system. The structure of the vacuum field
equations is analyzed and the exact solutions for the field
variables in the metric tensor are found in the form of expansions
in powers of a radial coordinate. We find that in the NGT axially
symmetric case the mass of the system remains constant only if the
system is static (as it necessarily is in the case of {\em spherical}
symmetry). If the system radiates, then the mass decreases monotonically
and the energy flux associated with waves is positive.

\vspace{.25in}

{\bf Revised version, June 1992}

\vspace{1in}
\begin{center}
{\large UTPT-92-01, gr-qc/9207009 }
\end{center}

\end{titlepage}

\section{Introduction}

The present work examines gravitational radiation in the
Nonsymmetric Gravitational Theory (NGT) (for a recent detailed
review see \cite{Moff91}). We probe the NGT asymptotic behaviour in
the wave zone using an {\em exact} solution.

In General Relativity (GR) gravitational radiation from bounded
sources has been studied not only through the linearized theory but
also with the use of exact solutions. The latter was done for the
general case of a bounded source in asymptotically flat spacetime
\cite{Sachs}. It was found that confining the arguments to the
axially symmetric case did not cause any essential loss of
generality. Since even the relevant GR calculations are very
tedious and the level of technical difficulty in the case of NGT
increases considerably, we limit ourselves to the axi--symmetric
case. The GR gravitational waves from isolated axially symmetric
reflexion--symmetric systems were studied in detail in \cite{BBM}.
Since our treatment of the axi--symmetric NGT case is rather
parallel, familiarity with this analysis is strongly recommended.

Since the NGT was introduced \cite{Moff79-1} there have been few
analytic solutions of the field equations published. The exact
solutions known to date include the spherically symmetric vacuum
case \cite{Moff79-2}, the spherically symmetric interior case
\cite{Sav89,Sav90} and Bianchi type I cosmological solutions with
and without matter \cite{Kunst79,Kunst80}.

This, at least in part, follows from the fact that deriving NGT
field equations relevant for particular cases of interest is not as
technically simple as may be suggested by its superficial
similarity to the corresponding GR situations. Firstly, since the
underlying geometry is non-Riemannian, neither the fundamental
metric tensor \(g_{\mu \nu}\) nor the affine connection is
symmetric. This does not constitute a serious problem for the
choice of the form of \(g_{\mu \nu}\), since we can always assume
that its nonsymmetric part takes on the isometries of the symmetric
part, which in turn has a well defined GR limit. On the other hand,
calculating the connection coefficients proves to be a tedious and
time consuming exercise, independently of the method chosen.
Secondly, the resultant formulae for the nonsymmetric connection
coefficients are extremely complicated for all but the simplest
forms of the metric, thus becoming unwieldy to use in the
derivation of --- still more complicated --- field equations.

The NGT quantities presented in this paper were derived with the
use of symbolic algebraic computation procedures. To this end we
have used the symbolic computation system {\em Maple}.

In Section \ref{fielde} we briefly summarize the necessary
fundamentals of NGT. Section \ref{coord} deals with a coordinate
system and generalization of the GR metric to the NGT case. Then in
Section \ref{res} we expand the metric in negative powers of a
suitably chosen radial coordinate and analyze the field equations.
The closing section contains conclusions.

Due to their unwieldiness, we retain most of the formulae in the
appendices and present the actual calculations only if their
tediousness is not forbidding. However, all the results and intermediate
steps of the calculations can be made available to the interested
reader. In the case of extremely unwieldy quantities the most
advisable form of doing that would be the transfer of computer files
in the internal {\em Maple} format.

Throughout this paper we use units in which \(G=c=1\).

\section{NGT Vacuum Field Equations} \label{fielde}

The NGT Lagrangian without sources takes the form:
\begin{equation} \label{lagr}
{\cal L} =  \sqrt{-g}g^{\mu \nu} R_{\mu \nu}(W),
\end{equation}
with \(g\) the determinant of \(g_{\mu \nu}\). The NGT Ricci tensor
is defined as:
\begin{equation} \label{ricciw}
R_{\mu \nu}(W) = W^{\beta}_{\mu \nu , \beta}- \frac{1}{2}
(W^{\beta}_{\mu \beta , \nu}+W^{\beta}_{\nu \beta , \mu}) -
W^{\beta}_{\alpha \nu}W^{\alpha}_{\mu \beta}+W^{\beta}_{\alpha
\beta}W^{\alpha}_{\mu \nu},
\end{equation}
and \(W^{\lambda}_{\mu \nu}\) is an unconstrained nonsymmetric
connection :
\begin{equation} \label{connw}
W^{\lambda}_{\mu \nu}=W^{\lambda}_{(\mu \nu)}+W^{\lambda}_{[\mu
\nu]}.
\end{equation}
(Throughout this paper parentheses and square brackets enclosing
indices stand for symmetrization and antisymmetrization,
respectively.)
The contravariant nonsymmetric tensor \(g^{\mu \nu}\) is defined in
terms of the equation:
\begin{equation} \label{inverse}
g^{\mu \nu} g_{\sigma \nu}=g^{\nu \mu} g_{\nu
\sigma}=\delta^{\mu}_{\sigma}.
\end{equation}
If we define the torsion vector as:
\begin{equation}
W_{\mu} \equiv W^{\nu}_{[\mu \nu]} = \frac{1}{2} \left(
W^{\nu}_{\mu \nu}- W^{\nu}_{\nu \mu} \right),
\end{equation}
then the connection:
\begin{equation} \label{conng}
\Gamma^{\lambda}_{\mu \nu} = W^{\lambda}_{\mu \nu} + \frac{2}{3}
\delta^{\lambda}_{\mu} W_{\nu}
\end{equation}
is torsion free:
\begin{equation} \label{gtors}
\Gamma_{\mu} \equiv \Gamma^{\alpha}_{[\mu \alpha]} = 0.
\end{equation}
Defining now:
\begin{equation} \label{riccig}
R_{\mu \nu}(\Gamma) = \Gamma^{\beta}_{\mu\nu,\beta} -
\frac{1}{2}(\Gamma^{\beta}_{(\mu\beta),\nu} + \Gamma^{\beta}_{(\nu
\beta) , \mu})-\Gamma^{\beta}_{\alpha\nu}\Gamma^{\alpha}_{\mu
\beta}+\Gamma^{\beta}_{(\alpha\beta)}\Gamma^{\alpha}_{\mu\nu},
\end{equation}
we can write:
\begin{equation} \label{ricciw=g}
R_{\mu \nu}(W) = R_{\mu \nu}(\Gamma) + \frac{2}{3} W_{[\mu ,\nu]},
\end{equation}
where \(W_{[\mu,\nu]}=\frac{1}{2}(W_{\mu,\nu}-W_{\nu,\mu})\).
Finally, the NGT vacuum field equations can be expressed as:
\begin{subequations} \label{fens}
\begin{equation} \label{fensgamma}
g_{\mu\nu,\sigma} - g_{\rho\nu} {\Gamma}^{\rho}_{\mu\sigma} -
g_{\mu\rho} {\Gamma}^{\rho}_{\sigma\nu} = 0 ,
\end{equation}
\begin{equation} \label{fensdiver}
{(\sqrt{-g}g^{[\mu \nu]})}_{ , \nu} = 0 ,
\end{equation}
\begin{equation} \label{fensricci}
R_{\mu \nu}(\Gamma) = \frac{2}{3} W_{[\nu , \mu]}.
\end{equation}
\end{subequations}
For the purpose of the analysis of Section \ref{res}, it is
convenient to decompose \(R_{\mu\nu}\) into standard symmetric and
antisymmetric parts: \(R_{(\mu\nu)}\), \(R_{[\mu\nu]}\), and then
rewrite the field equation (\ref{fensricci}) in the following form:
\begin{subequations} \label{fensriccis}
\begin{equation} \label{sym}
R_{(\mu \nu)}(\Gamma) = 0,
\end{equation}
\begin{equation} \label{asym}
R_{[\mu \nu , \rho]}(\Gamma) = 0,
\end{equation}
\end{subequations}
where we used equations (\ref{conng}), (\ref{gtors}) and the
notation:
\begin{equation}
R_{[\mu\nu,\rho]} = {R_{[\mu\nu]}}_{,\rho} + {R_{[\nu\rho]}}_{,\mu}
+ {R_{[\rho\mu]}}_{,\nu} .
\end{equation}

\section{The Metric} \label{coord}

Similarly to GR, the simplest NGT field due to a bounded source
would be spherically symmetric. However, the NGT equivalent of
Birkhoff's theorem (see e.g. \cite{Moff91}) shows that a
spherically symmetric gravitational field in an empty space must be
static. Hence, no gravitational radiation escapes into empty space
from a pulsating spherically symmetric source.

Following \cite{BBM} we consider the next simplest case: the field
which was initially static and spherically symmetric and eventually
becomes such, but undergoes an intermediate non--spherical wave
emitting period. Also, spacetime is assumed to be axially symmetric
and reflexion--symmetric at all times. Even now, due to the
complexity of the field equations, we are forced to use the method
of expansion to examine the problem. This approach, namely
expanding in negative powers of a radial coordinate, was also used
in the GR analysis \cite{BBM} and seems to naturally suit a wave
problem.

Due to the physical picture sketched above and to the fact that we
are interested in the asymptotic behaviour of the field at spatial
infinity (in an arbitrary direction from our isolated source) polar
coordinates \(x^{0}=u, {\bf x} = (r,\theta,\phi)\) are the natural
choice. The ``retarded time'' \(u\) has the property that the
hypersurfaces \(u=\mbox{const.}\) are light--like. Detailed
discussion of the coordinate systems permissible for investigation
of outgoing gravitational waves from isolated systems can be found
in \cite{Sachs,BBM}.

The covariant GR metric tensor corresponding to the situation
described above:
\begin{equation} \label{grmetric}
^{GR}g_{\mu\nu} = \left( \begin{array}{cccc}
Vr^{-1}e^{2\beta}-U^{2}r^{2}e^{2\gamma} & e^{2\beta} &
Ur^{2}e^{2\gamma} & 0 \\
e^{2\beta} & 0 & 0 & 0 \\
Ur^{2}e^{2\gamma} & 0 & -r^{2}e^{2\gamma} & 0 \\
0 & 0 & 0 & -r^{2}e^{-2\gamma}\sin^{2}\theta \end{array} \right),
\end{equation}
with \(U,V,\beta,\gamma\) being functions of \(u,r\) and
\(\theta\) was first given in \cite{Bondi60}.

For any metric in polar coordinates form conditions must be imposed
in the neighbourhood of the polar axis, \(\sin\theta=0\), to ensure
regularity. In the case under consideration we have that, as
\(\sin\theta\rightarrow0\),
\begin{equation} \label{condition}
V,\beta,U/\sin\theta,\gamma/\sin^{2}\theta
\end{equation}
each is a function of \(\cos\theta\) regular at \( \cos\theta = \pm
1 \).

The NGT generalization of the metric tensor (\ref{grmetric}) is:
\begin{equation} \label{genmetric}
g_{\mu\nu} = \left( \begin{array}{cccc}
Vr^{-1}e^{2\beta}-U^{2}r^{2}e^{2\gamma} & e^{2\beta} + \omega &
Ur^{2}e^{2\gamma} + \lambda & 0 \\
e^{2\beta} - \omega & 0 & 0 & 0 \\
Ur^{2}e^{2\gamma} - \lambda & 0 & -r^{2}e^{2\gamma} & 0 \\
0 & 0 & 0 & -r^{2}e^{-2\gamma}\sin^{2}\theta \end{array} \right),
\end{equation}
where \(\omega\) and \(\lambda\) are functions of \(u,r\) and
\(\theta\), and the field equation (\ref{fensdiver}) imposes the
following conditions:
\begin{subequations} \label{ngtcond}
\begin{eqnarray}
\label{cond0}
\mu=0 &:& 2\omega^{3} - e^{4\beta} \left[ 2(1 - 2r\beta_{,r})\omega
+ r\omega_{,r} \right] = 0 , \\  \nonumber \\
\label{cond1}
\pagebreak
\mu=1 &:& e^{2\beta} \left[ (\cot\theta - 2\gamma_{,\theta})\lambda
+ \lambda_{,\theta} \right] \nonumber \\& & + r^{2} e^{2\gamma}
\left[ (2\beta_{,u} + 2U\beta_{,\theta} - U_{,\theta} -U\cot\theta)
\omega -\omega_{,u} -\omega_{,\theta} \right] \nonumber \\& & +
e^{2\beta} \left[ 2(\gamma_{,\theta}-\beta_{,\theta}) \omega^{2}
\lambda - \omega^{2} (\lambda_{,\theta} + \lambda\cot\theta) +
\omega\omega_{,\theta}\lambda \right] \nonumber \\& & + r^{2} e^{-
2(\beta-\gamma)} \left( U_{,\theta} + U\cot\theta \right)
\omega^{3} = 0 , \\
\label{cond2}
\mu=2 &:& e^{2\beta} (\lambda_{,r} - 2\gamma_{,r}\lambda) -r
e^{2\gamma} \left[ (2U -2rU\beta_{,r} + rU_{,r})\omega + rU
\omega_{,r} \right] \nonumber \\& &+e^{-2\beta} \left[ 2(
\gamma_{,r}-\beta_{,r})\omega^{2}\lambda - \omega^{2}\lambda_{,r}
+ \omega\omega_{,r} \right] \nonumber \\& &+ re^{-2(\beta-
\gamma)}(U+rU_{,r}) \omega^{3} = 0.
\end{eqnarray}
\end{subequations}
We first solve the equation (\ref{cond0}) for \(\omega\), then substitute
the solution into (\ref{cond2}) and solve it for \(\lambda\). The solutions
are:
\begin{subequations} \label{omlambda}
\begin{equation} \label{omega}
\omega=e^{2\beta}{\left(1+C_{\omega}r^{4}\right)}^{-1/2},
\end{equation}
\begin{equation} \label{lambda}
\lambda=e^{2\gamma}(U+C_{\lambda})r^{2}{\left(1+C_{\omega}r^4\right)}^{-1/2},
\end{equation}
\end{subequations}
where \(C_{\omega}(u,\theta)\) and \(C_{\lambda}(u,\theta)\) are functions of
integration satisfying the condition following from the equation (\ref{cond1}):
\begin{equation} \label{mucond}
C_{\omega,u} - C_{\lambda}C_{\omega,\theta} + 2C_{\omega}\left(
C_{\lambda,\theta}+C_{\lambda}\cot\theta \right)=0.
\end{equation}
It may be noted here that the functions \(\omega\) and \(\lambda\) must both
be present in the metric (\ref{genmetric}) to ensure its regularity. The
case \(\omega\neq0\),\(\lambda=0\) leads to a singular \(g^{\mu\nu}\), while
the case \(\omega=0\),\(\lambda\neq0\) leads to a solution for \(\lambda\)
singular on the polar axis.

Since we consider an asymptotically flat spacetime, it follows from
(\ref{lambda}) that to satisfy the boundary conditions we must have
\(C_{\lambda}=0\). This in turn, due to the condition (\ref{mucond}),
restricts \(C_{\omega}\) to be a constant. To maintain a notation consistent
with other NGT results, we write this constant as:
\[
C_{\omega}=l^{-4}.
\]
The skew components of the metric are now:
\begin{subequations} \label{skews}
\begin{equation} \label{lomega}
\omega=\frac{e^{2\beta}l^{2}}{{(l^{4}+r^{4})}^{1/2}},
\end{equation}
\begin{equation} \label{llambda}
\lambda=\frac{e^{2\gamma}Ur^{2}l^{2}}{{(l^{4}+r^{4})}^{1/2}}.
\end{equation}
\end{subequations}

Finally, the NGT covariant metric tensor we work with takes the form:
\begin{equation} \label{covmetric}
g_{\mu\nu} = \left( \begin{array}{cccc}
Vr^{-1}e^{2\beta}-U^{2}r^{2}e^{2\gamma} & e^{2\beta}\left(1+
\frac{l^{2}}{{(l^{4}+r^{4})}^{1/2}} \right) & e^{2\gamma}Ur^{2} \left(
1+ \frac{l^{2}}{{(l^{4}+r^{4})}^{1/2}} \right) & 0 \\
e^{2\beta} \left(1-\frac{l^{2}}{{(l^{4}+r^{4})}^{1/2}} \right) & 0 & 0 & 0 \\
e^{2\gamma}Ur^{2} \left(1- \frac{l^{2}}{{(l^{4}+r^{4})}^{1/2}} \right) & 0 &
-r^{2}e^{2\gamma} & 0 \\
0 & 0 & 0 & -r^{2}e^{-2\gamma}\sin^{2}\theta \end{array} \right).
\end{equation}
The contravariant metric tensor obtained through (\ref{inverse}) is given by:
\begin{equation} \label{contrmetric}
g^{\mu\nu} = \left( \begin{array}{cccc}
0 & e^{-2\beta}{\left(1+\frac{l^{2}}{{(l^{4}+r^{4})}^{1/2}}\right)}^{-1}
& 0 & 0 \\
e^{-2\beta}{\left(1-\frac{l^{2}}{{(l^{4}+r^{4})}^{1/2}}\right)}^{-1}
& -Vr^{-1}e^{-2\beta}\frac{l^{4}+r^{4}}{r^{4}}+U^{2}e^{-4\beta}e^{2\gamma}
\frac{l^{4}}{r^{2}} & e^{-2\beta} U & 0 \\
0 & e^{-2\beta} U & -r^{-2} e^{-2\gamma} & 0 \\
0 & 0 & 0 & -r^{-2}e^{2\gamma}\sin^{-2}\theta \end{array} \right).
\end{equation}

\section{The Field Equations} \label{res}

Affine connection components for the metric (\ref{covmetric}) are
obtained by solving the system of 64 equations (\ref{fensgamma}).
The list of non--zero components is given in Appendix \ref{aconn}.
The Ricci tensor is then calculated with the use of (\ref{riccig})
and the standard decomposition of its non--zero components into
symmetric and antisymmetric parts is performed. The symmetric components
\(R_{(\mu\nu)}\) relevant to our calculation can be found in Appendix
\ref{aricci}. For the sake of convenience in comparing our results
to the GR case \cite{BBM}, we have split each \(R_{(\mu\nu)}\) into
the GR part \(^{GR}R_{\mu\nu}\) and the NGT contributions.

The only non--zero antisymmetric Ricci tensor components for the case
at hand are \(R_{[01]},R_{[02]},R_{[12]}\). However, the antisymmetric
field equation (\ref{asym}):
\begin{equation} \label{anti}
R_{[01,2]}=R_{[01],2}+R_{[12],0}+R_{[20],1}=0 ,
\end{equation}
as will be demonstrated shortly, must be trivially satisfied as a
consequence of the NGT Bianchi identities. For this reason we do not
present these skew components (except for their expanded forms) in this
paper.

The seven non--zero symmetric components of the Ricci tensor belong
to three categories. The symmetric field equations (\ref{sym}) for
four of them:
\begin{equation} \label{main}
R_{11}=R_{(12)}=R_{22}=R_{33}=0 ,
\end{equation}
will be called  ---following the terminology of \cite{BBM}--- ``the main
equations''. The other two non--trivial symmetric equations:
\begin{equation} \label{suppl}
R_{00}=R_{(02)}=0 ,
\end{equation}
we will call ---again following \cite{BBM}--- ``the supplementary
conditions''. The remaining symmetric field equation
\[
R_{(01)}=0
\]
is, similarily to the equation (\ref{anti}), identically satisfied due to the
Bianchi identities.

The generalized Bianchi identities of NGT can be written \cite{Al} (also
\cite{kunmoff}) in the form:
\begin{equation} \label{BI1}
\sqrt{-g}g^{\mu\alpha} \left( R_{\mu+\alpha-;\rho} - R_{\mu+\rho+;\alpha}
-R_{\rho-\alpha-;\mu} \right)=0,
\end{equation}
where we used Einstein's notation:
\begin{subequations}
\begin{eqnarray}
{A^{\beta+}}_{;\alpha} &=& {A^{\beta}}_{,\alpha} + A^{\epsilon}
\Gamma^{\beta}_{\epsilon\alpha}, \\
{A^{\beta-}}_{;\alpha} &=& {A^{\beta}}_{,\alpha} + A^{\epsilon}
\Gamma^{\beta}_{\alpha\epsilon}.
\end{eqnarray}
\end{subequations}
We rewrite (\ref{BI1}) in the form:
\begin{equation} \label{BI2}
\sqrt{-g}g^{\mu\alpha} \left( R_{\mu\alpha,\rho} - R_{\mu\rho,\alpha}
-R_{\rho\alpha,\mu} +2 R_{(\rho\lambda)} \Gamma^{\lambda}_{\mu\alpha}
\right)=0.
\end{equation}
If we now recall the form of the metric (\ref{contrmetric}) and suppose
that the four main equations (\ref{main}) are satisfied, then the Bianchi
identities reduce to:
\begin{subequations} \label{FBI}
\begin{eqnarray}
\label{FBI0}
\rho=0 &:& g^{\mu\alpha}\Gamma^{2}_{\mu\alpha}R_{(02)} -g^{(01)}R_{00,1}
        -g^{11}R_{(01),1} -g^{22}R_{(02),2} \nonumber \\
       & & +g^{[12]}(R_{(01),2} + R_{(02),1}) +
        g^{(12)}(R_{[01],2}+R_{[12],0}+R_{[20],1})=0, \\
\label{FBI1}
\rho=1 &:& g^{\mu\alpha}\Gamma^{0}_{\mu\alpha}R_{(01)}=0, \\
\label{FBI2}
\rho=2 &:& g^{\mu\alpha}\Gamma^{0}_{\mu\alpha}R_{(02)}
        +g^{[01]}(R_{(01),2} + R_{(02),1}) \nonumber \\
       & & +g^{(01)}(R_{[01],2}+R_{[12],0}+R_{[20],1})=0 .
\end{eqnarray}
\end{subequations}
 From (\ref{FBI1}), we see that \(R_{(01)}\) vanishes as a consequence of
the main equations. Also, if supplementary conditions are satisfied, then
the only non--trivial antisymmetric field equation (\ref{anti}) must hold.

We now rewrite the main equations in the form:
\begin{subequations} \label{mn}
\begin{eqnarray}
\label{mn1}
\hspace{-1in}0=R_{11}&=& \frac{4r^{4}}{l^{4}+r^{4}} \left[ \left(
\frac{{\gamma_{,r}}^{2}}{2} - \frac{\beta_{,r}}{r} \right) -
\frac{l^{4}}{r^{2}(l^{4}+r^{4})} \right], \\
\label{mn2}
0=-2r^{2} R_{(12)}&=& {\left( r^{4} e^{2(\gamma-\beta)} U_{,r}\right)}_{,r}
\nonumber \\ & & -2 r^2 \left[ 2\gamma_{,r}(\gamma_{,\theta} - \cot\theta) -
\frac{2\beta_{,\theta}}{r} + \beta_{,r\theta} - \gamma_{,r\theta}
\right] \\& & - \frac{2l^{4}}{l^{4}+r^{4}}e^{2(\gamma-\beta)}r
\left[3U(1+2r{\beta}_{,r})-rU_{,r}-2r^{2}U{\gamma}_{,r}(1+r {\gamma}_{,r})
\right], \nonumber \\
\label{mn3}
0=R_{22}e^{2(\beta-\gamma)} - r^{2}R^{3}_{3}e^{2\beta}&=&2V_{,r} +
\frac{r^{4}}{2}e^{-2(\beta-\gamma)}{U_{,r}}^{2} - r^{2}
{(U\cot\theta+U_{,\theta})}_{,r} \nonumber \\& &- 4r
(U\cot\theta+U_{,\theta}) -2 e^{2(\beta-\gamma)} \left[1 +
(3\gamma_{,\theta} -\beta_{,\theta})\cot\theta \right. \nonumber
\\& &+ \left. 2\gamma_{,\theta} (\beta_{,\theta} -
\gamma_{,\theta}) -\beta_{,\theta\theta} +\gamma_{,\theta\theta} -
{\beta_{,\theta}}^{2} \right] \\& & -\frac{2l^{4}}{l^{4}+r^{4}}
e^{2(\gamma-\beta)}r^{2}\left[U^{2}(1+2r\beta_{,r}-r^{2}{{\gamma}_{,r}}^{2})
+rUU_{,r} \right] \nonumber \\& & + \frac{2l^{4}}{l^{4}+r^{4}} \left[
r(U\cot\theta+U_{,\theta})-\frac{2V}{r} \right], \nonumber \\
\label{mn4}
0=- r^{2}R^{3}_{3}e^{2\beta}&=&2r{(r\gamma)}_{,ur} + (1-
r\gamma_{,r})(V_{,r}-rU_{,\theta}) - (r\gamma_{,rr}+\gamma_{,r})V
\nonumber \\& &-2 e^{2(\beta-\gamma)} \left[1 + (3\gamma_{,\theta}
-2\beta_{,\theta})\cot\theta + 2\gamma_{,\theta} (\beta_{,\theta} -
 \gamma_{,\theta}) +\gamma_{,\theta\theta} \right] \nonumber \\& &
+ r \left[ 2{(r\gamma_{,r} + \gamma)}_{,\theta} +(r\gamma_{,r}-
3)\cot\theta \right]U \nonumber \\& &+r^{2}(\gamma_{,\theta}-
\cot\theta)U_{,r} \\& &- \frac{2l^{4}}{l^{4}+r^{4}}
e^{-2\beta}\frac{1}{r}\left[\gamma_{,u}+U({\gamma}_{,r}-\cot\theta)
+\frac{V}{r^{2}} (1-r\gamma_{,r}) \right]. \nonumber
\end{eqnarray}
\end{subequations}

The structure of the above equations is analogous to the GR case.
The first three equations include only derivatives on the
hypersurface \(u=\mbox{const}\). The last equation,
however, contains differentiation with respect to \(u\). Thus, if
\(\gamma\) is given for some value of \(u\), then (\ref{mn1})
determines \(\beta\). In turn (\ref{mn2}) gives \(U\) and then
\(V\) follows from (\ref{mn3}). Having all those functions,
one can use (\ref{mn4}) to find \(\gamma\) at the later instant of \(u\).
One can then repeat the whole cycle. Thus, if \(\gamma\) is given for
some value of \(u\), the equations (\ref{mn}) determine the temporal
evolution of our system except for functions of integration. The latter
are arbitrary functions of \(u\) and \(\theta\), but are independent of \(r\).

The equations (\ref{mn}) are independent. This is to be expected, since
the number of independent NGT field equations must correspond to the
number of independent field variables, when the NGT Bianchi identities
are taken into account (see e.g. \cite{moffmath}). It may also be noted
that the system (\ref{mn}) reduces to its GR counterpart of ref. \cite{BBM}
on substitution \(l^2=0\).

The analysis leading to determining the form of the functions
\(U,V,\beta,\gamma\) in our case duplicates the one given in detail
in \cite{BBM}. The reason is straightforward. Upon inspection of the
equations, we find that if the solutions \(U,V,\beta,\gamma\) have forms
of an expansion in powers of \(r\), then the NGT terms of the equations
do not contribute to the leading terms. Further, the functions
of integration in (\ref{mn}) must remain unchanged, if we are to
recover the GR limit upon setting \(l^{2}=0\). This can be easiest
seen in the case of the equation (\ref{mn1}). If \(\gamma\) were to be
the same in GR and NGT cases, then the NGT solution for \(\beta\) would
have the form:
\[
\beta_{NGT}=\frac{1}{4}\ln\left(C_{\beta}(u,\theta)+\frac{l^{4}}{r^{4}}\right)
+\beta_{GR}= \frac{1}{4}\ln\left(1+\frac{l^{4}}{r^{4}}\right)+\beta_{GR},
\]
where the second equation takes into account the boundary condition.
The leading term in the expansion of \(\beta_{GR}\) is of order \(r^{-2}\)
and the highest order NGT contribution goes like \(r^{-4}\). Therefore,
any NGT effect must fall off faster than the GR effects in the wave
zone.

The reader interested in the details of determining the functions
\(U,V,\beta,\gamma\) is referred to \cite{BBM}. Here we only
summarize the results.

The requirement that the field contain only outgoing radiation at
large distances from the source gives the form of \(\gamma\):
\[
\gamma=\frac{f(t-r)}{r}+\frac{g(t-r)}{r^{2}}+...
\]
Substituting this into (\ref{mn}), isolating the functions of
integration and applying suitable coordinate transformations, gives
the following:
\begin{subequations} \label{expansions}
\begin{eqnarray} \label{expbeta}
\beta=&-&\frac{1}{4}c^{2}r^{-2}+...\nonumber \\
      &+&\frac{1}{4}l^{4}r^{-4}-\frac{1}{8}l^{8}r^{-8}+...,
\end{eqnarray}
\begin{equation} \label{expgamma}
\gamma=cr^{-1}+\left(C-\frac{1}{6}c^{3}\right)r^{-3}+... ,
\end{equation}
\begin{eqnarray} \label{expU}
U=&-&(c_{,\theta}+2c\cot\theta)r^{-2} + (2N + 3cc_{,\theta} +
4c^{2}\cot\theta) r^{-3} \nonumber \\&+& \left( \frac{3}{2}
C_{,\theta} +3C\cot\theta -3cN -4c^{2}c_{,\theta} -4c^{3}\cot\theta
\right) r^{-4} + ... \nonumber \\
&-&\frac{2}{3}(c_{,\theta}+2c\cot\theta)l^{4}r^{-6}+...,
\end{eqnarray}
\begin{eqnarray} \label{expV}
V=&r& -2M - \left[N_{,\theta} +N\cot\theta -{c_{,\theta}}^{2} -
4cc_{,\theta}\cot\theta -\frac{1}{2}c^{2}(1+8\cot^{2}\theta)
\right] r^{-1} \nonumber \\&-& \frac{1}{2} \left[ C_{,\theta\theta}
+3C_{,\theta}\cot\theta -2C +6N(c_{,\theta}+2c\cot\theta) \right.
\nonumber \\& & +\left. 8c({c_{,\theta}}^{2} +3cc_{,\theta}
+2c^{2}\cot^{2}\theta) \right] r^{-2} + ... \nonumber \\&+& \frac{1}{2}
l^{4}r^{-3} - \frac{1}{6}\left(6M + \frac{1}{4}c_{,\theta\theta}
+\frac{3}{4}c_{,\theta}\cot\theta + c \right)l^{4}r^{-4} + ... ,
\end{eqnarray}
\end{subequations}
where \(c(u,\theta),N(u,\theta),M(u,\theta)\) are the functions of
integration and \(C(u,\theta)\) satisfies:
\[
4C_{,u}=2c^{2}c_{,u}+2cM+N\cot\theta-N_{,\theta}.
\]
In (\ref{expansions}), we isolated the NGT contributions from the rest of
the expansion. Clearly, the leading terms are purely GR ones.

One can readily verify the expansions (\ref{expansions}) by a
direct substitution into (\ref{mn}). The expanded non--zero affine
connection components are given in the Appendix \ref{aexpconn}.
Again, for the sake of convenience of comparison to the GR results,
we split the expansions into the NGT and GR parts. We draw the
reader's attention to certain errors occurring in the expanded
forms of \(\Gamma^{0}_{22}\) and \(\Gamma^{1}_{12}\), as published
in \cite{BBM}.

The expansions (\ref{expansions}) considerably simplify the form of
the supplementary conditions (\ref{suppl}). These extremely complex
equations (see Appendix \ref{aricci}) reduce now to the leading
inverse square--law terms involving only the relations of the
functions \(c,M,N\) and the lower order terms including pure NGT
contributions and the coupling of the two. We see that, as we
anticipated earlier from the form of the equations (\ref{mn}), the
NGT skew contributions decay rapidly with increasing distance from the
source and the dominant asymptotic behaviour of the system at spatial
infinity is given by \(r^{-2}\) terms. From the latter, we get:
\begin{subequations} \label{SSS}
\begin{eqnarray}
\label{massderiv}
M_{,u}&=& -{c_{,u}}^{2} + \frac{1}{2} {(c_{,\theta\theta}
+3c_{,\theta}\cot\theta -2c)}_{,u}   , \\
\label{Nderiv}
-3N_{,u}&=& M_{,\theta} +3cc_{,u\theta} +4cc_{,u}\cot\theta
+c_{,u}c_{,\theta}   ,
\end{eqnarray}
\end{subequations}
in full agreement with the GR results. We note that the equations
(\ref{SSS}) are {\em exact }, if the series expansions are valid.

To complete the discussion of the asymptotic behaviour of the NGT
quantities, we expand the only non--zero components of \(R_{[\mu
\nu]}\) :
\begin{subequations} \label{expskewriccis}
\begin{eqnarray}
R_{[01]}&=& l^{2} \left\{\left(4M + 4cc_{,u} +c_{,\theta\theta}+
3c_{,\theta}\cot\theta-2c\right) r^{-5}+ ... \right\}, \\R_{[02]}&=&
l^{2} \left\{3{(c_{,\theta}+2c\cot\theta)}_{,u} r^{-3} \right.
\\& & \left. + \left[4c_{,u}(c\cot\theta-c_{,\theta}) +
3c(c_{,\theta}+2c\cot\theta) +3c_{,\theta}-M_{,\theta} \right] r^{-4}
+ ... \right\}, \nonumber \\R_{[12]}&=& l^{2} \left\{3(c_{,\theta}
+2c\cot\theta) r^{-4} -2\left[c(3c_{,\theta} +2c\cot\theta)
+3N\right]r^{-5} + ... \right\}.
\end{eqnarray}
\end{subequations}
 From (\ref{fensricci}), it now follows that the torsion vector
\(W_{\mu}\) has the maximum asymptotic behaviour \(1/r^{3}\). This
clearly contradicts the result of \cite{DDM} obtained through a
perturbation expansion of the NGT field equations about a purely
Einstein local vacuum background. The reader is referred to \cite{CorMoff}
for a discussion of the inconsistencies of the latter result.

In order to identify \(M\) and \(N\) one has to turn to the static
metrics. We once again refer the reader interested in the details
to \cite{BBM}. After connecting the metric (\ref{covmetric}) to the
static axially symmetric metric in the Weyl form (through a lengthy
transformation), one finds that in the static case \(M(u,\theta)\)
reduces to the mass \(m\) of the system and is independent of both
its arguments. The functions \(N\) and \(C\) are related to the
dipole and quadrupole moments, respectively. Also, \(c\) is an
arbitrary function of \(\theta\) in the static case. In the
time--dependent case the entire information about the temporal
evolution of the system is contained in \(c(u,\theta)\).

Since, in this paper, we are interested only in the physical
interpretation of the function \(M\) (\(N\) and \(C\) can be
examined with the use of the above mentioned transformation), we
limit our analysis here to the following simple exercise.
First let us write the non--zero components of the metric tensor
(\ref{covmetric}) using the expansions (\ref{expansions}):
\begin{subequations} \label{expmetric}
\begin{eqnarray}
\label{expmetric00}
g_{00}&=&1-\frac{2M}{r}-\frac{1}{r^{2}}(N_{,\theta}+N\cot\theta)+ ...
\nonumber \\& &+\frac{l^{4}}{r^{4}} -\frac{l^{4}}{r^{5}} \left( 2M
+\frac{1}{8}c_{,\theta}\cot\theta +\frac{1}{6}c +\frac{1}{24}
c_{,\theta\theta}\right)+ ... ,\\
\label{expmetric01s}
g_{(01)}&=&1-\frac{c^{2}}{2r^{2}}+\frac{1}{2r^{4}}\left(3Cc-
\frac{c^{4}}{2} \right)+ \frac{l^{4}}{2r^{4}} + ... , \\
\label{expmetric01as}
g_{[01]}&=&\frac{l^{2}}{r^{2}} - \frac{c^{2}l^{2}}{2r^{4}} + ... , \\
\label{expmetric02s}
g_{(02)}&=&-c_{,\theta} -2c\cot\theta + \frac{2N+cc_{,\theta}}{r}+
... , \\
\label{expmetric02as}
g_{[02]}&=& -\frac{l^{2}}{r^{2}}\left(c_{,\theta}+2c\cot\theta\right)
+\frac{l^{2}}{r^{3}}\left(2N+cc_{,\theta}\right)+ ... , \\
\label{expmetric22}
g_{22}&=& -r^{2} -c^{2} -2cr -\frac{1}{r} \left( 2C -
\frac{1}{3}c^{3} \right) + ... , \\
\label{expmetric33}
g_{33}&=& \sin^{2}\theta \left[-r^{2} -c^{2} +2cr +\frac{1}{r}
\left( 2C - \frac{1}{3}c^{3} \right) + ... \right] .
\end{eqnarray}
\end{subequations}

We confine ourselves to initial and final static systems.
Suppose that the system is in a dynamic period, which can be
physically interpreted only as the emission of waves, between the
moments \(u_{i}\) and \(u_{f}\). Thus, for \(u \leq u_{i}\) and \(u
\geq u_{f}\) the function \(c\) does not vary. The static limit
\(c(u,\theta) \rightarrow c_{s}(\theta)\) of (\ref{expmetric}) gives
the equivalent of the static Weyl metric in our case. However, to
demonstrate the physical interpretation of \(M\), we simplify the
situation further. We can scale the function \(c\) for either one of the
static periods to be \(c=0\) (forsaking the \(\theta\) dependence
of \(c\) limits us here to a static spherically symmetric system).
We now remove the terms containing the functions \(N\) and \(C\) from
(\ref{expmetric}). We have a {\em post factum} justification for doing
that from the interpretation of \(N\) and \(C\) as multipole moments.
Since there is no radiation during the static period, then \(N=C=0\).
The metric (\ref{expmetric}) tends now to its static spherically
symmetric limit:
\begin{subequations} \label{expstmetric}
\begin{eqnarray}
\label{expstmetric00}
g_{00}&=&1-\frac{2M_{s}}{r} +\frac{l^{4}}{r^{4}} -
\frac{2M_{s}l^{4}}{r^{5}} , \\
\label{expstmetric01s}
g_{(01)}&=&1+\frac{l^{4}}{2r^{4}} , \\
\label{expstmetric01as}
g_{[01]}&=&\frac{l^{2}}{r^{2}} , \\
\label{expstmetric02}
g_{(02)}&=&g_{[02]}=0 ,  \\
\label{expstmetric22}
g_{22}&=& -r^{2} , \\
\label{expstmetric33}
g_{33}&=& -r^{2} \sin^{2}\theta  ,
\end{eqnarray}
\end{subequations}
where by \(M_{s}\) we denote the static limit of \(M\).

Now a coordinate transformation from our retarded time \(u\) to the
usual time coordinate \(t=u+r\) converts (\ref{expstmetric})
into the NGT static spherically symmetric metric \cite{Moff91,Moff79-2}:
\begin{equation} \label{NGTSchwarz}
{ds}^2 = (1 + \frac{l^4}{r^4})(1 - \frac{2M_{s}}{r}){dt}^2 - {(1 -
\frac{2M_{s}}{r})}^{-1}{dr}^2 - r^2 \left( {d\theta}^2 +
{\sin}^2\theta{d\phi}^2 \right),
\end{equation}
with
\[
 g_{[01]} = \frac{{l^2}}{{r^2}}.
\]
Thus, the static spherically symmetric limit \(M_{s}\) of the
``mass aspect'' \(M(u,\theta)\) can only be interpreted as the mass
of the system.

If we define the mass \(m(u)\) of the system as the mean value of
\(M(u,\theta)\) over the sphere:
\begin{equation} \label{mass}
m(u)=\frac{1}{2} \int^{\pi}_{0} M(u,\theta)\sin\theta d\theta ,
\end{equation}
then \(c(u,\theta)\) completely determines the time evolution of
the mass \(m(u)\). Integrating (\ref{massderiv}) and noticing that
the second term in it does not contribute to the integral due to
the conditions (\ref{condition}), we get:
\begin{equation} \label{centralresult}
m_{,u}=\frac{dm}{du}=-\frac{1}{2} \int^{\pi}_{0} {c_{,u}}^{2}
\sin\theta d\theta .
\end{equation}
Since we discussed here systems whose initial and final states are
static, the physical interpretation of \(m(u)\) as the mass of the
system is unambiguous. Analogously to the GR case the main result
is as follows:

{\em The mass of an axially symmetric NGT system is constant only
if the system remains static. If the system evolves in time (emits
waves), the mass decreases monotonically.}

Since radiation is the only energy loss mechanism available to the
system, the above proves that gravitational waves emitted by an
axi--symmetric reflexion symmetric NGT source compatible with the
metric (\ref{covmetric}) carry positive energy or, in other words,
the flux of gravitational energy in NGT is positive.

\section{Conclusions} \label{con}

We have proved that an NGT axi--symmetric system emitting
gravitational waves has the usual GR-like asymptotic behaviour in
the wave zone. The NGT contributions to the physical quantities
decay rapidly with the distance from the source and the energy flux
at spatial infinity is necessarily positive.

Whether a similar result could be obtained in the general NGT case
without any symmetries remains to be verified, although it is expected
that a proof of this would be similar to the one in \cite{Sachs}.

\vspace{.5in}
{\bf Acknowledgements}

This work was supported by the Natural Sciences and Engineering
Research Council of Canada. We are grateful to N. Cornish and J. Spencer
for helpful discussions.

\newpage
\appendix

\section{List of non--zero affine connection components}
\label{aconn}

\begin{eqnarray}
\Gamma^{0}_{00}&=&^{GR}\Gamma^{0}_{00} + \frac{4l^{4}}{r^{2}(l^{4}+r^{4})}
\left(U^{2}r^{3}e^{-2(\beta-\gamma)}-V\right), \nonumber \\
^{GR}\Gamma^{0}_{00}&=&2 \beta_{,u} + r^{2} e^{2(\beta-\gamma)} U
\left( U_{,r} + \frac{U}{r} + \gamma_{,r}U \right) -
\frac{\beta_{,r}V}{r} - \frac{V_{,r}}{2r} + \frac{V}{2r^{3}},\\
\Gamma^{0}_{01}&=& - \frac{2l^{2}}{r(l^{4}+r^{4})}\left(l^{2}-
\sqrt{l^{4}+r^{4}}\right), \\
\Gamma^{0}_{02}&=&^{GR}\Gamma^{0}_{02} - \frac{l^{2}r}{l^{4}+r^{4}}
\left(3l^{2}+\sqrt{l^{4}+r^{4}}\right)Ue^{-2(\beta-\gamma)},
\nonumber \\
^{GR}\Gamma^{0}_{02}&=&\beta_{,\theta} - r^{2} e^{-2(\beta-\gamma)}
\left(\frac{U_{,r}}{2} + \frac{U}{r} + \gamma_{,r}U \right),\\
\Gamma^{0}_{10}&=& - \frac{2l^{2}}{r(l^{4}+r^{4})}\left(l^{2}+
\sqrt{l^{4}+r^{4}}\right), \\
\Gamma^{0}_{20}&=&^{GR}\Gamma^{0}_{02} - \frac{l^{2}r}{l^{4}+r^{4}}
\left(3l^{2}-\sqrt{l^{4}+r^{4}}\right)Ue^{-2(\beta-\gamma)},\\
\Gamma^{0}_{22}&=&^{GR}\Gamma^{0}_{22}=r e^{-2(\beta-\gamma)}
\left( 1+r\gamma_{,r} \right),\\
\Gamma^{0}_{33}&=&^{GR}\Gamma^{0}_{33}=r e^{-2(\beta + \gamma)}
\sin^{2}\theta \left( 1-r\gamma_{,r} \right),\\
\Gamma^{1}_{00}&=&^{GR}\Gamma^{1}_{00} + \frac{4l^{4}}{r^{3}(l^{4}+r^{4})}
V^{2}+\frac{l^{4}}{r(2l^{4}+r^{4})}UV_{,\theta}
+\frac{2l^{4}r^{3}}{l^{4}+r^{4}}U^{4} e^{-4(\beta-\gamma)} \nonumber \\
& &+\frac{l^{4}r^{4}}{2l^{4}+r^{4}}U^{3}U_{,r} e^{-4(\beta-\gamma)}
 -\frac{3l^{4}(3l^{4}+r^{4})}{(l^{4}+r^{4})
(2l^{4}+r^{4})}VU^{2} e^{-2(\beta-\gamma)} \nonumber \\
& & +\frac{rl^{4}} {2l^{4}+r^{4}} e^{-2(\beta-\gamma)}\left[2VU^{2}
{(\gamma-\beta)}_{,r}+ U(U_{,r}V- UV_{,r}) \right. \nonumber \\
& & \hspace{1.3cm} \left. +2rU^{3}{(\beta-\gamma)}_{,\theta} + 4rU^{2}
{(\beta-\gamma)}_{,u} +2rU(U_{,u}+UU_{,\theta}) \right], \nonumber \\
^{GR}\Gamma^{1}_{00}&=&\frac{V_{,u}}{2r} - \frac{\beta_{,u}V}{r} -
\frac{U V_{,\theta}}{2r} -\frac{\beta_{,\theta}UV}{r}
+\frac{VV_{,r}}{2r^{2}} +\frac{\beta_{,r}V^{2}}{r^{2}} -
\frac{V^{2}}{2r^{3}} \nonumber \\
& & + r^{2} e^{-2(\beta-\gamma)} \left[ U^{2} \left(U_{,\theta} +
\gamma_{,\theta}U - \frac{V}{r^{2}} - \frac{\gamma_{,r}V}{r} +
\gamma_{,u} \right) - \frac{UU_{,r}V}{r} \right],\\
\Gamma^{1}_{01}&=&^{GR}\Gamma^{1}_{01} +\frac{2l^{4}}{l^{4}+r^{4}}
\left[\frac{2V}{r^{2}} - rU^{2}e^{-2(\beta-\gamma)} \right]
 +\frac{l^{2}}{\sqrt{l^{4}+r^{4}}}\left(U{\beta}_{,\theta}
-\frac{2V}{r^{2}}\right) \nonumber \\& &-\frac{l^{2}}{\sqrt{l^{4}+r^{4}}}
r e^{-2(\beta-\gamma)} \left[\frac{l^{4}}{l^{4}+r^{4}}U^{2} +rU^{2}
\gamma_{,r} +\frac{1}{2}rUU_{,r} \right] , \nonumber \\
^{GR}\Gamma^{1}_{01}&=&\frac{V_{,r}}{2r} - \frac{V}{2r^{2}}
+\frac{\beta_{,r}V}{r} - \beta_{,\theta}U - \frac{1}{2}r^{2} e^{-
2(\beta-\gamma)} UU_{,r},\\
\Gamma^{1}_{02}&=&^{GR}\Gamma^{1}_{02} -\frac{l^{4}}{2l^{4}+r^{4}}
V_{,\theta} - \frac{l^{4}}{2l^{4}+r^{4}}\left( \frac{2}{l^{4}+r^{4}}
r^{7}U^{3} +\frac{1}{2}U^{2}U_{,r} \right) e^{-4(\beta-\gamma)} \nonumber
\\& & +\frac{l^{4}}{2(2l^{4}+r^{4})} e^{-2(\beta-\gamma)}\left\{
UV[2r{(\beta-\gamma)}_{,r}-3]+r(UV_{,r}-U_{,r}V) \right. \nonumber \\
& &\hspace{3.8cm} \left. +2r^{2}[U^{2}{(\gamma-\beta)}_{,\theta}+U_{,u}
+UU_{,\theta}-2U{(\beta-\gamma)}_{,u}] \right\} \nonumber \\& & +
\frac{l^{4}}{(l^{4}+r^{4})(2l^{4}+r^{4})}r^{3}UV e^{-2(\beta-\gamma)}
-\frac{l^{2}\sqrt{l^{4}+r^{4}}}{2l^{4}+r^{4}}V_{,\theta} \nonumber \\
& &+ \frac{l^{6}}{(2l^{4}+r^{4})\sqrt{l^{4}+r^{4}}}r^{4}U^{2} \left(
rU_{,r}+2U\right)e^{-4(\beta-\gamma)} \nonumber \\& &+ \frac{l^{2}
\sqrt{l^{4}+r^{4}}}{2l^{4}+r^{4}}e^{-2(\beta-\gamma)} \left\{ 2r^{3}
[U(U_{,\theta}+U\gamma_{,\theta})+U{(\gamma-2\beta)}_{,u}+U_{,u}]
\right. \nonumber \\& &\hspace{3.8cm} \left.+r^{2}(UV_{,r}-U_{,r}V)
+2r^{2}UV\beta_{,r}-rUV \right\} \nonumber \\& &+ \frac{l^{6}}
{(2l^{4}+r^{4})\sqrt{l^{4}+r^{4}}}e^{-2(\beta-\gamma)} \left[ 2U(V+
r^{2}V_{,r})-2r^{3}U(U\beta_{,\theta}-\gamma_{,u}) \right] \nonumber
\\& &+\frac{2l^{2}}{\sqrt{l^{4}+r^{4}}}rUVe^{-2(\beta-\gamma)}
+\frac{4l^{2}{(l^{4}+r^{4})}^{3/2}}{r(2l^{4}+r^{4})}U^{2}
\beta_{,\theta}e^{-2(\beta-\gamma)} , \nonumber \\
^{GR}\Gamma^{1}_{02}&=&\frac{V_{,\theta}}{2r} + r^{2} e^{-2(\beta-
\gamma)} \left[U \left( \frac{V}{r^{2}} + \frac{\gamma_{,r}V}{r} -
\gamma_{,u} -U_{,\theta} -\gamma_{,\theta}U \right) +
\frac{U_{,r}V}{2r} \right], \\
\Gamma^{1}_{10}&=&^{GR}\Gamma^{1}_{01} +\frac{2l^{4}}{l^{4}+r^{4}}
\left[\frac{2V}{r^{2}} - rU^{2}e^{-2(\beta-\gamma)} \right]
 -\frac{l^{2}}{\sqrt{l^{4}+r^{4}}}\left(U{\beta}_{,\theta}
-\frac{2V}{r^{2}}\right) \nonumber \\& &+\frac{l^{2}}{\sqrt{l^{4}+r^{4}}}
r e^{-2(\beta-\gamma)} \left[\frac{l^{4}}{l^{4}+r^{4}}U^{2} +rU^{2}
\gamma_{,r} +\frac{1}{2}rUU_{,r} \right] , \\
\Gamma^{1}_{11}&=&^{GR}\Gamma^{1}_{11}+\frac{4l^{4}}{(l^{4}+r^{4})r},
\nonumber \\
^{GR}\Gamma^{1}_{11}&=&2\beta_{,r},\\
\Gamma^{1}_{12}&=&^{GR}\Gamma^{1}_{12} + e^{-2(\beta-\gamma)}Ur \left[
\frac{l^{2}}{\sqrt{l^{4}+r^{4}}}r\gamma_{,r}-\frac{3l^{4}}{l^{4}+r^{4}}
\right] , \nonumber \\
^{GR}\Gamma^{1}_{12}&=&\beta_{,\theta} + \frac{1}{2}r^{2} e^{-
2(\beta-\gamma)} U_{,r}, \\
\Gamma^{1}_{20}&=&^{GR}\Gamma^{1}_{02} -\frac{l^{4}}{2l^{4}+r^{4}}
V_{,\theta} - \frac{l^{4}}{2l^{4}+r^{4}}\left( \frac{2}{l^{4}+r^{4}}
r^{7}U^{3} +\frac{1}{2}U^{2}U_{,r} \right) e^{-4(\beta-\gamma)} \nonumber
\\& & +\frac{l^{4}}{2(2l^{4}+r^{4})} e^{-2(\beta-\gamma)}\left\{
UV[2r{(\beta-\gamma)}_{,r}-3]+r(UV_{,r}-U_{,r}V) \right. \nonumber \\
& &\hspace{3.8cm} \left. +2r^{2}[U^{2}{(\gamma-\beta)}_{,\theta}+U_{,u}
+UU_{,\theta}-2U{(\beta-\gamma)}_{,u}] \right\} \nonumber \\& & +
\frac{l^{4}}{(l^{4}+r^{4})(2l^{4}+r^{4})}r^{3}UV e^{-2(\beta-\gamma)}
+\frac{l^{2}\sqrt{l^{4}+r^{4}}}{2l^{4}+r^{4}}V_{,\theta} \nonumber \\
& &- \frac{l^{6}}{(2l^{4}+r^{4})\sqrt{l^{4}+r^{4}}}r^{4}U^{2} \left(
rU_{,r}+2U\right)e^{-4(\beta-\gamma)} \nonumber \\& &- \frac{l^{2}
\sqrt{l^{4}+r^{4}}}{2l^{4}+r^{4}}e^{-2(\beta-\gamma)} \left\{ 2r^{3}
[U(U_{,\theta}+U\gamma_{,\theta})+U{(\gamma-2\beta)}_{,u}+U_{,u}]
\right. \nonumber \\& &\hspace{3.8cm} \left.+r^{2}(UV_{,r}-U_{,r}V)
+2r^{2}UV\beta_{,r}-rUV \right\} \nonumber \\& &- \frac{l^{6}}
{(2l^{4}+r^{4})\sqrt{l^{4}+r^{4}}}e^{-2(\beta-\gamma)} \left[ 2U(V+
r^{2}V_{,r})-2r^{3}U(U\beta_{,\theta}-\gamma_{,u}) \right] \nonumber
\\& &-\frac{2l^{2}}{\sqrt{l^{4}+r^{4}}}rUVe^{-2(\beta-\gamma)}
-\frac{4l^{2}{(l^{4}+r^{4})}^{3/2}}{r(2l^{4}+r^{4})}U^{2}
\beta_{,\theta}e^{-2(\beta-\gamma)} , \\
\Gamma^{1}_{21}&=&^{GR}\Gamma^{1}_{12} - e^{-2(\beta-\gamma)}Ur \left[
\frac{l^{2}}{\sqrt{l^{4}+r^{4}}}r\gamma_{,r}-\frac{3l^{4}}{l^{4}+r^{4}}
\right] ,  \\
\Gamma^{1}_{2 2}&=&^{GR}\Gamma^{1}_{2 2} + \frac{l^{4}}{l^{4}+r^{4}}
U^{2}r^{3}e^{-4(\beta- \gamma)} , \nonumber \\
^{GR}\Gamma^{1}_{2 2}&=&r^{2} e^{-2(\beta-\gamma)} \left(
\gamma_{,u} + U_{,\theta} +\gamma_{,\theta}U - \frac{V}{r^{2}} -
\frac{\gamma_{,r}V}{r} \right), \\
\Gamma^{1}_{33}&=&^{GR}\Gamma^{1}_{33}=r^{2} \sin^{2}\theta e^{-
2(\beta+\gamma)} \left( -\gamma_{,u} + U\cot\theta -
\gamma_{,\theta}U - \frac{V}{r^{2}} + \frac{\gamma_{,r}V}{r}
\right), \\
\Gamma^{2}_{00}&=&^{GR}\Gamma^{2}_{00} -\frac{l^{4}}{l^{4}+r^{4}}
\frac{2UV}{r^{2}} -\frac{l^{4}}{2l^{4}+r^{4}}\frac{V_{,\theta}}{r^{3}}
e^{2(\beta-\gamma)} \nonumber \\& &+\left[ \frac{l^{4}}{l^{4}+r^{4}}
rU^{3}-\frac{l^{4}}{2l^{4}+r^{4}}r^{2}U^{2}U_{,r}\right]
e^{-2(\beta-\gamma)} \nonumber \\& &-\frac{l^4}{(2l^{4}+r^{4})r^2}
\left[3UV-r(UV_{,r}-U_{,r}V)-2rUV{(\beta-\gamma)}_{,r} \right.
\nonumber \\& &\hspace{2.8cm}\left. +4r^2U{(\beta-\gamma)}_{,u}
+2r^2U^{2}{(\beta-\gamma)}_{,\theta} -2r^2(U_{,u}+UU_{,\theta})\right]
, \nonumber \\
^{GR}\Gamma^{2}_{00}&=&-U_{,u} + U \left( 2\beta_{,u} -2\gamma_{,u}
-U_{,\theta}- \gamma_{,\theta}U + \frac{V}{2r^{2}} -
\frac{V_{,r}}{2r} - \frac{\beta_{,r}V}{r} \right) \nonumber \\& &
+e^{2(\beta-\gamma)} \frac{V_{,\theta} + 2\beta_{,\theta}
V}{2r^{3}} +r^{2} e^{-2(\beta-\gamma)} U^{2} \left( U_{,r} +
\frac{U}{r} +\gamma_{,r}U \right), \\
\Gamma^{2}_{01}&=&^{GR}\Gamma^{2}_{01} +\frac{l^2}{\sqrt{l^4+r^4}}
\frac{2\beta_{,\theta}}{r^2}e^{2(\beta-\gamma)} -
\frac{l^2}{\sqrt{l^4+r^4}}\left[\frac{2U}{r}(1+r\gamma_{,r})+U_{,r}
\right] \nonumber \\& &-\frac{l^2}{\sqrt{l^4+r^4}}{\left(
\frac{l^2}{\sqrt{l^4+r^4}}-1\right)}^2\frac{2U}{r} , \nonumber \\
^{GR}\Gamma^{2}_{01}&=& - \frac{U_{,r}}{2} -\frac{U}{r} -\gamma_{,r}U +
\frac{\beta_{,\theta}}{r^{2}} e^{2(\beta-\gamma)},\\
\Gamma^{2}_{02}&=&^{GR}\Gamma^{2}_{02} -\frac{l^2}{\sqrt{l^4+r^4}}
\left[\frac{V\gamma_{,r}}{r}-\gamma_{,u}-U\beta_{,\theta}+V\right]
\nonumber \\& &+\frac{l^2}{2l^4+r^4}r^2UU_{,r}e^{-2(\beta-\gamma)}
\nonumber \\& &-\frac{l^2}{\sqrt{l^4+r^4}}{\left(\frac{l^2}
{\sqrt{l^4+r^4}}+1\right)}^2rU^2e^{-2(\beta-\gamma)}, \nonumber \\
^{GR}\Gamma^{2}_{02}&=&\gamma_{,u} + \beta_{,\theta}U -r^{2} e^{-
2(\beta-\gamma)} U \left( \frac{U_{,r}}{2} + \frac{U}{r}
+\gamma_{,r}U \right), \\
\Gamma^{2}_{10}&=&^{GR}\Gamma^{2}_{01} -\frac{l^2}{\sqrt{l^4+r^4}}
\frac{2\beta_{,\theta}}{r^2}e^{2(\beta-\gamma)} +
\frac{l^2}{\sqrt{l^4+r^4}}\left[\frac{2U}{r}(1+r\gamma_{,r})+U_{,r}
\right] \nonumber \\& &+\frac{l^2}{\sqrt{l^4+r^4}}{\left(
\frac{l^2}{\sqrt{l^4+r^4}}-1\right)}^2\frac{2U}{r} ,  \\
\Gamma^{2}_{12}&=&^{GR}\Gamma^{2}_{12}-\frac{l^2}{\sqrt{l^4+r^4}}
\left(\frac{1}{r}+\gamma_{,r}\right), \nonumber \\
^{GR}\Gamma^{2}_{12}&=&\frac{1}{r} + \gamma_{,r}, \\
\Gamma^{2}_{20}&=&^{GR}\Gamma^{2}_{02} +\frac{l^2}{\sqrt{l^4+r^4}}
\left[\frac{V\gamma_{,r}}{r}-\gamma_{,u}-U\beta_{,\theta}+V\right]
\nonumber \\& &-\frac{l^2}{2l^4+r^4}r^2UU_{,r}e^{-2(\beta-\gamma)}
\nonumber \\& &+\frac{l^2}{\sqrt{l^4+r^4}}{\left(\frac{l^2}{
\sqrt{l^4+r^4}}-1\right)}^2rU^2e^{-2(\beta-\gamma)}, \\
\Gamma^{2}_{21}&=&^{GR}\Gamma^{2}_{12}+\frac{l^2}{\sqrt{l^4+r^4}}
\left(\frac{1}{r}+\gamma_{,r}\right), \\
\Gamma^{2}_{22}&=&^{GR}\Gamma^{2}_{2 2}=\gamma_{,\theta} + r^{2}
e^{-2(\beta-\gamma)} U \left( \frac{1}{r} +\gamma_{,r} \right), \\
\Gamma^{2}_{33}&=&^{GR}\Gamma^{2}_{33}=r^{2} \sin^{2}\theta
e^{2(\beta-\gamma)} U \left( \frac{1}{r} -\gamma_{,r} \right)
\nonumber \\ & & -e^{-4\gamma} \sin^{2}\theta \left(\cot\theta -
\gamma_{,\theta} \right),\\
\Gamma^{3}_{03}&=&^{GR}\Gamma^{3}_{03} -\frac{l^2}{\sqrt{l^4+r^4}}
\left[\gamma_{,u}+\left(\frac{1}{r}-\gamma_{,r}\right)\left(
\frac{V}{r}-r^2U^2e^{-2(\beta-\gamma)}\right)\right], \nonumber \\
^{GR}\Gamma^{3}_{03}&=&-\gamma_{,u}, \\
\Gamma^{3}_{13}&=&^{GR}\Gamma^{3}_{13}-\frac{l^2}{\sqrt{l^4+r^4}}
\left(\frac{1}{r} - \gamma_{,r}\right), \nonumber \\
^{GR}\Gamma^{3}_{13}&=&\frac{1}{r} - \gamma_{,r}, \\
\Gamma^{3}_{23}&=&^{GR}\Gamma^{3}_{23} -\frac{l^2}{\sqrt{l^4+r^4}}
\left( \frac{1}{r} -\gamma_{,r} \right)r^2U e^{-2(\beta-\gamma)},
\nonumber \\ ^{GR}\Gamma^{3}_{23}&=&\cot\theta-\gamma_{,\theta}, \\
\Gamma^{3}_{30}&=&^{GR}\Gamma^{3}_{03} +\frac{l^2}{\sqrt{l^4+r^4}}
\left[\gamma_{,u}+\left(\frac{1}{r}-\gamma_{,r}\right)\left(
\frac{V}{r}-r^2U^2e^{-2(\beta-\gamma)}\right)\right], \\
\Gamma^{3}_{31}&=&^{GR}\Gamma^{3}_{13}+\frac{l^2}{\sqrt{l^4+r^4}}
\left(\frac{1}{r} - \gamma_{,r}\right), \\
\Gamma^{3}_{32}&=&^{GR}\Gamma^{3}_{23} +\frac{l^2}{\sqrt{l^4+r^4}}
\left( \frac{1}{r} -\gamma_{,r} \right)r^2U e^{-2(\beta-\gamma)}.
\end{eqnarray}

\newpage
\section{List of non--zero symmetric Ricci tensor components} \label{aricci}

\begin{eqnarray}
\hspace{-1in}R_{00}=^{GR}R_{00} &-& \frac{l^4}{2l^4+r^4}
e^{2(\beta-\gamma)}\frac{1}{r^3}\left[2V_{,\theta}{(\beta-\gamma)}_
{,\theta}+V_{,\theta}\cot\theta+V_{,\theta\theta}\right] \nonumber \\
&+&\frac{l^4}{r(2l^4+r^4)}\left\{2UV{[{(\beta-\gamma)}_{,\theta}
+(\beta-\gamma)\cot\theta]}_{,r}-4r(U_{,\theta}+U\cot\theta)
{(\beta-\gamma)}_{,u} \right. \nonumber \\& &\hspace{2.1cm}-2r
(2UU_{,\theta}+U^2\cot\theta){(\beta-\gamma)}_{,\theta}+
2r(U_{,u\theta}+{U_{,\theta}}^2+UU_{,\theta\theta} \nonumber \\& &
\hspace{2.1cm}+2(U_{,\theta}V+2UV_{,\theta}){(\beta-\gamma)}_{,r}+
(UV_{,r}-U_{,r}V)\cot\theta \nonumber \\& &\hspace{2.1cm} +2r
(U_{,u}+UU_{,\theta})\cot\theta-4rU{(\beta-\gamma)}_{,u\theta}
\nonumber \\ & &\hspace{2.1cm}\left.-2rU^2{(\beta-\gamma)}_
{,\theta\theta}-{(U_{,r}V)}_{,\theta}+U_{,\theta}V_{,r}+
2UV_{,r\theta} \right\} \nonumber \\&-&\frac{l^4}{{(l^4+r^4)}^2}
\frac{l^4+9r^4}{r^4}V^2- \frac{l^4}{(l^4+r^4)(2l^4+r^4)}
\frac{7l^4+5r^4}{r^2}(U\cot\theta+U_{,\theta})V \nonumber \\
&-& \frac{l^4}{(l^4+r^4){(2l^4+r^4)}^2} \frac{8l^8+14l^4r^4+7r^8}{r^2}
UV_{,\theta} \nonumber \\ &-&\frac{l^4}{r^2(l^4+r^4)}
\left[V_{,u}-2r^2{\gamma_{,u}}^2+2V(2r\gamma_{,u}\gamma_{,r}-
\beta_{,u}-V{\gamma_{,r}}^2+U\beta_{,\theta})\right. \nonumber
\\& &\hspace{2.2cm} \left.+\frac{V(2V\beta_{,r} -3V_{,r})}{r}\right]
\nonumber \\&+& \frac{l^4}{2l^4+r^4}r^2Ue^{-2(\beta-\gamma)}
\left\{6UU_{,r}{(\beta-\gamma)}_{,\theta}+4{\left[U{(\beta-\gamma)}_
{,u}\right]}_{,r}+2U^2{(\beta-\gamma)}_{,r\theta} \right.\nonumber \\
& &\hspace{3.8cm} \left. -U{(3U_{,\theta}+U\cot\theta)}_{,r} -
2(U_{,ur}+2U_{,r}U_{,\theta}) \right\} \nonumber \\&-& \frac{l^4}
{2l^4+r^4}rUe^{-2(\beta-\gamma)}\left\{UV_{,rr}-U_{,rr}V+
{\left[UV{(\beta-\gamma)}_{,r}\right]}_{,r}\right\} \nonumber \\
&+& \frac{2l^4}{l^4+r^4}rU^2e^{-2(\beta-\gamma)}\left(2r\gamma_{,u}
\gamma_{,r}+U\cot\theta-2V{\gamma_{,r}}^2\right)
\nonumber \\&+&\frac{2l^4}{{(l^4+r^4)}^2{(2l^4+r^4)}^2}\left(
r^{12}+10l^{12}+17l^4r^8+29l^8r^4\right)\frac{V}{r}U^2e^{-2(\beta-\gamma)}
\nonumber \\&-&\frac{2l^4}{{(l^4+r^4)}^2{(2l^4+r^4)}^2}\left(
l^{12}-3r^{12}+l^4r^8+8l^8r^4\right)r\gamma_{,u}U^2e^{-2(\beta-\gamma)}
\nonumber \\&+& \frac{2l^4}{(l^4+r^4){(2l^4+r^4)}^2}
\left[2r^9(UU_{,\theta}+U_{,u}-2U\beta_{,u})-r(4l^8+4l^4r^4-r^8)
U^2\gamma_{,\theta}\right. \nonumber \\& & \hspace{3.7cm} -
(4l^8+2l^4r^4-r^8)UV_{,r}+(10l^8+11l^4r^4+5r^8)UV\beta_{,r} \nonumber
\\& & \hspace{3.7cm} \left. +l^4(2l^4+3r^4)U_{,r}V+(2l^8+l^4r^4-2r^8)
UV\gamma_{,r} \right]Ue^{-2(\beta-\gamma)} \nonumber \\
&+&\frac{16l^8}{{(2l^4+r^4)}^2}rU^3\beta_{,\theta}e^{-2(\beta-\gamma)}
- \frac{2l^4}{l^4+r^4}r^3U^4e^{-4(\beta-\gamma)}\left(2\beta_{,r}
-r{\gamma_{,r}}^2\right) \nonumber \\&+& \left[ \frac{2l^4}
{{(2l^4+r^4)}^2}(6l^4+r^4)r^3U_{,r}-\frac{2l^4(2l^4+r^4)}
{{(l^4+r^4)}^2}rU\right] U^3e^{-4(\beta-\gamma)} \nonumber \\&+&
\frac{l^4}{2l^4+r^4}r^4U^2e^{-4(\beta-\gamma)}\left[UU_{,rr}+
2{U_{,r}}^2 -2UU_{,r}{(\beta-\gamma)}_{,r}\right] , \\
^{GR}R_{00}=&-&2U(\beta_{,u\theta} - \gamma_{,u\theta}) -
2U_{,\theta}(\beta_{,u} - \gamma_{,u}) + U_{,u\theta} +
UU_{,\theta\theta} + {U_{,\theta}}^{2} \nonumber \\&-&
2UU_{,\theta} ( \beta_{,\theta} - \gamma_{,\theta}) + U^{2} (2
{\beta_{,\theta}}^{2} -2\beta_{,\theta}\gamma_{,\theta} +
\gamma_{,\theta\theta}) +2{\gamma_{,u}}^{2} \nonumber \\&-&
\cot\theta \left[ 2U(\beta_{,u} - \gamma_{,u}) -U_{,u} -
UU_{,\theta} - U^{2}\gamma_{,\theta} \right] \nonumber \\&+& e^{-
2(\beta-\gamma)} \left[ 4UU_{,r}V + U^{2}(V\gamma_{,r} + V_{,r})
\right] - \frac{1}{2r} (U_{,r}V_{,\theta} - U_{,\theta}V_{,r} -
UV_{,r}\cot\theta) \nonumber \\&+& \frac{1}{r} \left[ 2V
(\beta_{,ur} + U\beta_{,r\theta}) + 2UV_{,\theta}( \beta_{,r} -
\gamma_{,r}) + U_{,\theta}V\beta_{,r} \right. \nonumber \\& &
\hspace{.4cm} \left.+ U(V_{,r}\beta_{,\theta} + V_{,r\theta} +
V\beta_{,r}\cot\theta) \right] -  \frac{1}{2r^{2}} V(U_{,\theta} +
V_{,rr} + U\cot\theta) \nonumber \\&-& \frac{1}{r^{2}} \left[
2U(V_{,\theta} -V\beta_{,\theta}) + V( V\beta_{,rr}
+V_{,r}\beta_{,r} -2\beta_{,u}) +V_{,u} \right] \nonumber \\&-&
\frac{1}{2r^{3}} e^{2(\beta-\gamma)} \left[ V_{,\theta\theta} +
2V\beta_{,\theta\theta} +\left(2\beta_{,\theta} - 2\gamma_{,\theta}
+\cot\theta \right) \left(V_{,\theta} + 2V\beta_{,\theta} \right)
\right] \nonumber \\&+& r e^{-2(\beta-\gamma)} \left[ UV (U_{,rr} -
 2 U_{,r}\beta_{,r} + 2 U_{,r}\gamma_{,r}) \right. \nonumber \\& &
\hspace{1.7cm}\left. +U^{2} \left(V\gamma_{,rr} + V_{,r} \gamma_{,r} -
3U_{,\theta} -2\gamma_{,u} \right) - U^{3} (\cot\theta + 2
\gamma_{,\theta}) + \frac{U_{,r}^{2}V}{2} \right] \nonumber \\&+&
r^{2} e^{-2(\beta-\gamma)} \left[-2U^{3}\gamma_{,r\theta} + U^{2}
\left( 2U_{,r} \beta_{,\theta} - 3U_{,r}\gamma_{,\theta} -
U_{,\theta}\gamma_{,r} -2\gamma_{,ur} -2U_{,r\theta} \right)
\right. \nonumber \\ & & \hspace{1.5cm}\left.-U^{2} (U_{,r}+U\gamma_{,r})
\cot\theta -U(U_{,ur} +2U_{,r}U_{,\theta}) + 2UU_{,r}(\beta_{,u} -
\gamma_{,u}) \right] \nonumber \\ &+& \frac{1}{2} r^{4} e^{-4(\beta-
\gamma)} U^{2}{U_{,r}}^{2},
\end{eqnarray}
\begin{eqnarray}
\hspace{-1in}R_{(01)}=^{GR}R_{01} &-& \frac{l^4}{l^4+r^4} \left[
\frac{2U_{,\theta}}{r}+\frac{2U}{r}\left(\beta_{,\theta}+\cot\theta
\right)-\frac{2V{\gamma_{,r}}^2}{r} +2\gamma_{,u}\gamma_{,r} +
\frac{2V\beta_{,r}- 3V_{,r}} {r^2}+\frac{V}{r^3} \right] \nonumber
\\&-&\frac{8l^4} {{(l^4+r^4)}^2} rV - \frac{2l^8}{{(l^4+r^4)}^2}U^2
e^{-2(\beta-\gamma)} \nonumber \\ &-& \frac{l^4}{l^4+r^4}
e^{-2(\beta-\gamma)}\left[2U^2(r^2 {\gamma_{,r}}^2-2r\beta_{,r}-1)
+ rUU_{,r} \right] , \\
^{GR}R_{01}=&-&U(\beta_{,r\theta} - \gamma_{,r\theta}) - \left(
\frac{U_{,r}}{2} + U\gamma_{,r} \right)\cot\theta -
U_{,r}\beta_{,\theta} - U_{,\theta}\gamma_{,r} -2\gamma_{,u}
\gamma_{,r} \nonumber \\&-& \frac{U_{,r\theta}}{2} - 2\beta_{,ur}
+ \frac{V\beta_{,r}}{r^{2}} + \frac{1}{r^{2}} e^{2(\beta-\gamma)}
\left( 2{\beta_{,\theta}}^{2} -2\beta_{,\theta}\gamma_{,\theta} +
\beta_{,\theta}\cot\theta + \beta_{,\theta\theta} \right) \nonumber
\\&-& \frac{1}{r} \left( 2U\beta_{,\theta} + U\cot\theta +
U_{,\theta} + V\beta_{,rr} - V_{,r}\beta_{,r} - \frac{V_{,rr}}{2}
\right) \nonumber \\&-& r^{2} e^{-2(\beta-\gamma)} \left[ UU_{,r}
\left( \frac{2}{r} - \beta_{,r} + \gamma_{,r} \right) +
\frac{UU_{,rr}}{2} + \frac{{U_{,r}}^{2}}{2} \right],
\end{eqnarray}
\begin{eqnarray}
\hspace{-1in}R_{(02)}=^{GR}R_{02} &+& \frac{4l^4}{{(2l^4+r^4)}^2}r^2
V_{,\theta} +\frac{l^4}{(2l^4+r^4)r}\left[V_{,\theta}(\gamma_{,r}-
2\beta_{,r}) - \frac{V_{,r\theta}}{2} \right] \nonumber \\
&-& \frac{2l^4}{l^4+r^4}r^3U^3\left[r{\gamma_{,r}}^2-2\beta_{,r}
\right]e^{-4(\beta-\gamma)} +\frac{2(2l^4+r^4)l^4}{{(l^4+r^4)}^2}
r^2U^3e^{-4(\beta-\gamma)} \nonumber \\ &-& \frac{l^4}{2l^4+r^4}r^4U
\left[UU_{,r}(\gamma_{,r}-\beta_{,r})+{U_{,r}}^2+\frac{UU_{,rr}}{2}
\right]e^{-4(\beta-\gamma)} \nonumber \\ &-& \left[\frac{4l^8}
{{(2l^4+r^4)}^2}+\frac{l^8}{(2l^4+r^4)(l^4+r^4)}\right]r^3U^2U_{,r}
e^{-4(\beta-\gamma)} \nonumber \\&-& \frac{2l^4}{l^4+r^4}e^{-2(\beta-
\gamma)} rU \left[r\gamma_{,u}\gamma_{,r}-V{\gamma_{,r}}^2+U\cot\theta
\right] \nonumber \\&-& \frac{l^4}{2l^4+r^4}r \left\{ \frac{U_{,rr}V-
UV_{,rr}}{2}-U_{,ur}-UV{(\beta-\gamma)}_{,rr} \right. \nonumber \\
& &\hspace{1.9cm}\left.+r{\left[U^2{(\beta-\gamma)}_{,\theta)}
\right]}_{,r}+2r{\left[U{(\beta-\gamma)}_{,u}\right]}_{,r} \right.
\nonumber \\& &\hspace{1.9cm} \left.-r(UU_{,r\theta}+U_{,r}U_{,\theta})
-(U_{,r}V+UV_{,r}){(\beta-\gamma)}_{,r} \right\} e^{-2(\beta-\gamma)}
\nonumber \\&+& \frac{l^4}{{(2l^4+r^4)}^2}e^{-2(\beta-\gamma)}
\left[ (2l^4+5r^4)UV\gamma_{,r}-(2l^4-r^4)U_{,r}V -4r^5U^2
\gamma_{,\theta}\right] \nonumber \\  \nonumber \\
&+& \frac{l^4} {(l^4+r^4) {(2l^4+r^4)}^2} \left[(4l^8-3r^8)
(2rU\gamma_{,u}+rU_{,u}-2rU\beta_{,u}) \right. \nonumber \\ & &
\hspace{3.5cm} \left. -(10l^8+15l^4r^4+7r^8)
UV\beta_{,r} +(8l^8+4l^4r^4-r^8)UV_{,r} \right. \nonumber \\& &
\hspace{3.5cm}\left.  -r(4l^8+6l^4r^4+3r^8)UU_{,\theta} \right]
e^{-2(\beta-\gamma)} \nonumber \\&-& \frac{l^8(13r^8-2l^8+19l^4r^4)}
{r(l^4+r^4) {(2l^4+r^4)}^2}UV ,\\
^{GR}R_{02}=& &\beta_{,u\theta}
-\gamma_{,u\theta} + 2\gamma_{,u}(\gamma_{,\theta} -\cot\theta) -
U(\beta_{,\theta\theta} + 2{\beta_{,\theta}}^{2} -2\beta_{,\theta}
\gamma_{,\theta} +\beta_{,\theta}\cot\theta) \nonumber \\&-& e^{-
2(\beta-\gamma)}(UV\gamma_{,r}+UV_{,r}+2U_{,r}V) -
\frac{V_{,r\theta}}{2r} -(\beta_{,r}-\gamma_{,r})
\frac{V_{,\theta}}{r} + \frac{V_{,\theta}}{2r^{2}} \nonumber \\&+&
r e^{-2(\beta-\gamma)} \left[ U\left( 3U_{,\theta} +2\gamma_{,r} -
V\gamma_{,rr} -V_{,r}\gamma_{,r} \right) +U^{2}(2\gamma_{,\theta}
+\cot\theta) \right. \nonumber \\&-& \left. \frac{U_{,rr}V}{2} +
U_{,r}V(\beta_{,r}-\gamma_{,r}) \right] + r^{2} e^{-2(\beta-
\gamma)} \left[ \frac{U_{,ur}}{2} + U_{,r}U_{,\theta} \right.
\nonumber \\&-& U_{,r}(\beta_{,u} -\gamma_{,u}) + U\left(
\frac{3}{2}U_{,r\theta} + 2\gamma_{,ur} + U_{,\theta}\gamma_{,r} +
\frac{U_{,r}}{2}\cot\theta \right) \nonumber \\&-& \left. UU_{,r}
(\beta_{,\theta}-\gamma_{,\theta}) + U^{2} ( 2\gamma_{,r\theta} +
\gamma_{,r}\cot\theta) \right] \nonumber \\&-& \frac{1}{2} r^{4}
e^{-4(\beta-\gamma)} U{U_{,r}}^{2},
\end{eqnarray}
\begin{eqnarray}
\hspace{-1in}R_{11}=^{GR}R_{11}&-&\frac{4l^4}{l^4+r^4} \left(
\frac{{\gamma_{,r}}^{2}}{2}-\frac{\beta_{,r}}{r}\right) -
\frac{4l^4r^2}{{(l^4+r^4)}^2} , \\
^{GR}R_{11}=&4&\left(\frac{{\gamma_{,r}}^{2}}{2}-\frac{\beta_{,r}}
{r}\right), \\
\hspace{-1in}R_{(12)}=^{GR}R_{12}&+&\frac{l^4}{l^4+r^4}e^{-2(\beta
-\gamma)}\left[2rU\gamma_{,r}(r\gamma_{,r}+1)-3U(2r\beta_{,r}+1)+
rU_{,r}\right], \\
^{GR}R_{12}=&2&\left[\gamma_{,r}(\cot\theta-
\gamma_{,\theta}) +\frac{\beta_{,\theta}}{r}\right] -
\beta_{,r\theta} + \gamma_{,r\theta} \nonumber \\&+& r^{2} e^{-
2(\beta-\gamma)} \left( U_{,r}\gamma_{,r} - U_{,r}\beta_{,r} +
\frac{2U_{,r}}{r} + \frac{U_{,rr}}{2} \right),
\end{eqnarray}
\begin{eqnarray}
\hspace{-1in}R_{22}=^{GR}R_{22} &-&\frac{6l^8}{{(l^4+r^4)}^2}
U^2r^2e^{-4(\beta-\gamma)} \nonumber \\&+& \frac{2l^4}{l^4+r^4}r^2
\left[U^2(r^2{\gamma_{,r}}^2-2r\beta_{,r}-1)-rUU_{,r}\right]
e^{-4(\beta-\gamma)} \nonumber \\ &+& \frac{2l^4}{l^4+r^4}
\left[r(U\gamma_{,\theta}+U_{,\theta}+\gamma_{,u}) -
V\gamma_{,r}-\frac{V}{r}\right]e^{-2(\beta-\gamma)} , \\
^{GR}R_{22}=& &1 + 2\left( \beta_{,\theta}\gamma_{,\theta} -
{\beta_{,\theta}}^{2} - {\gamma_{,\theta}}^{2} -
\beta_{,\theta\theta} + \frac{\gamma_{,\theta\theta}}{2} \right) +
3\gamma_{,\theta}\cot\theta - e^{-2(\beta-\gamma)} \left(
V\beta_{,r}+V_{,r} \right) \nonumber \\&+& r e^{-2(\beta-\gamma)}
\left( 2\gamma_{,u} +2U\gamma_{,\theta} +U\cot\theta +3U_{,\theta}
-V\gamma_{,rr} -V_{,r}\gamma_{,r} \right) \nonumber \\&+& r^{2}
e^{-2(\beta-\gamma)} \left( 2\gamma_{,ur} +U_{,\theta}\gamma_{,r}
+U\gamma_{,r}\cot\theta +U_{,r\theta} -2U\gamma_{,r\theta} \right)
\nonumber \\&+&r^{2}{U_{,r}}^{2}\gamma_{,\theta} - r^{4} e^{-
4(\beta-\gamma)} \frac{{U_{,r}}^{2}}{2}, \\
\hspace{-2in}R_{33}/\sin^{2}\theta=& &^{GR}R_{33}/\sin^{2}\theta
\nonumber \\&-&
\frac{2l^4}{l^4+r^4}\left[r(U\gamma_{,\theta}-U\cot\theta+\gamma_{,u})
-V\gamma_{,r}+\frac{V}{r}\right]e^{-2(\beta-\gamma)},\quad \\
\hspace{-2in}^{GR}R_{33}/\sin^{2}\theta=& &
e^{-2(\beta-\gamma)}(V\gamma_{,r}-V_{,r}) \nonumber \\&+& e^{-
4\gamma} \left[1+2\beta_{,\theta}(\gamma_{,\theta}-\cot\theta) +
3\gamma_{,\theta}\cot\theta +\gamma_{,\theta\theta} -
2{\gamma_{,\theta}}^{2} \right] \nonumber \\&+& r e^{-2(\beta-
\gamma)} \left( U_{,\theta} -2\gamma_{,u} -2U\gamma_{,\theta} +
3U\cot\theta +V\gamma_{,rr} +V_{,r}\gamma_{,r} \right) \nonumber
\\&+& r^{2} \left[ (U_{,r} -U\gamma_{,r})\cot\theta -
U_{,\theta}\gamma_{,r} - U_{,r}\gamma_{,\theta} -2\gamma_{,ur} -
2U\gamma_{,r\theta} \right].
\end{eqnarray}

\newpage
\section{Expanded non--zero affine connection components}
\label{aexpconn}

\begin{eqnarray}
\Gamma^{0}_{00}=&-& \frac{1}{r^{2}}(M+cc_{,u}) - \frac{1}{r^{3}}
(N\cot\theta+N_{,\theta}) + ... \nonumber \\&-& \frac{2l^4}{r^{5}}
+\frac{l^4}{2r^6}\left[7M-\frac{5}{2}\left(\frac{c}{3}+\frac{c_{,\theta}
\cot\theta}{4}+\frac{c_{,\theta\theta}}{12}\right)\right]+ ... ,\\
\Gamma^{0}_{01}=&-& \frac{2l^2}{r^{3}} - \frac{2l^4}{r^5}+ ...,\\
\Gamma^{0}_{02}=& & \frac{N}{r^{2}} + \frac{1}{r^{3}} \left(
\frac{3C_{,\theta}}{2} + Nc + 3C\cot\theta \right) + ... \nonumber
\\&+& \frac{l^2}{r^3}\left(c_{,\theta}+2c\cot\theta\right)-
\frac{l^2}{r^4}\left(2N+cc_{,\theta}\right)+ ...,\\
\Gamma^{0}_{10}=& & \frac{2l^2}{r^{3}} - \frac{2l^4}{r^5}+ ...,\\
\Gamma^{0}_{20}=& & \frac{N}{r^{2}} + \frac{1}{r^{3}} \left(
\frac{3C_{,\theta}}{2} + Nc + 3C\cot\theta \right) + ... \nonumber
\\&-& \frac{l^2}{r^3}\left(c_{,\theta}+2c\cot\theta\right)+
\frac{l^2}{r^4}\left(2N+cc_{,\theta}\right)+ ...,\\
\Gamma^{0}_{22}=& &c +r - \frac{c^{2}}{2r} - \frac{1}{r^{2}} \left(
\frac{c^{3}}{3} + C \right) + ... \nonumber \\&-& \frac{l^4}{2r^3}
-\frac{l^4}{2r^4}c+\frac{3l^4}{8r^5}c^2+ ...,\\
\Gamma^{0}_{33}/\sin^{2}\theta=& & r - c + \frac{c^{2}}{2r} -
\frac{1}{r^{2}} \left( \frac{c^{3}}{3} + C \right) + ... \nonumber \\
&-&\frac{l^4}{2r^3}+\frac{l^4}{2r^4}c+\frac{3l^4}{8r^5}c^2+ ..., \\
\Gamma^{1}_{00}=&-&\frac{M_{,u}}{r} + \frac{1}{r^{2}} \left[ M + c
c_{,u}(1+4\cot^{2}\theta) - \frac{1}{2} {(N\cot\theta +
N_{,\theta})}_{,u} \right. \nonumber \\&+& \left.
c_{,\theta}c_{,u\theta} + 2(c_{,u}c_{,\theta} +
cc_{,u\theta})\cot\theta \right] + ... \nonumber \\&+&
\frac{2l^2}{r^5}-\frac{l^4}{2r^5}\left[M_{,u}+\frac{1}{2}{\left(
\frac{c}{3}+\frac{c_{,\theta}\cot\theta}{4}+\frac{c_{,\theta\theta}}
{12}\right)}_{,u}\right]+ ...,\\
\Gamma^{1}_{01}=& & \frac{M}{r^{2}} + \frac{1}{r^{3}}(N\cot\theta
+ N_{,\theta}) + ... \nonumber \\&+& \frac{2l^2}{r^{3}}-
\frac{4l^2}{r^4}M+\frac{2l^4}{r^5}+ ... , \\
\Gamma^{1}_{02}=& & \frac{1}{r} \left[ c_{,u} ( c_{,\theta} +
2c\cot\theta) - M_{,\theta} \right] \nonumber \\&-&
\frac{1}{2r^{2}} \left[ 2c_{,u}(cc_{,\theta} + 2N) + N +
N_{,\theta\theta} + (N_{,\theta} - N\cot\theta)\cot\theta \right]
+ ... \nonumber \\&-& \frac{l^2}{r^2}{\left(c_{,\theta}+2c\cot\theta
\right)}_{,u} \\&+&\frac{l^2}{r^3}\left[cc_{,u\theta}
+2c\cot\theta(c_{,u} -2) +2c_{,\theta}(c_{,u}-1)+2N_{,u}+M_{,\theta}
\right]+ ..., \nonumber \\
\Gamma^{1}_{10}=& &  \frac{M}{r^{2}} + \frac{1}{r^{3}}(N\cot\theta
+ N_{,\theta}) + ... \nonumber \\&-& \frac{2l^2}{r^{3}}+
\frac{4l^2}{r^4}M+\frac{2l^4}{r^5}+ ... , \\
\Gamma^{1}_{11}=& & \frac{c^{2}}{r^{3}} + ... +\frac{2l^4}{r^5}
+ ...,\\
\Gamma^{1}_{12}=& & \frac{1}{r} (c_{,\theta}
+2c\cot\theta) - \frac{1}{r^{2}} (3N + 3cc_{,\theta} -2c^{2}
\cot\theta) + ... \nonumber \\ &+& \frac{l^2}{r^4}\left(cc_{,\theta}
+2c^2\cot\theta \right)-\frac{l^2}{r^5}c\left(cc_{,\theta}+2N\right)+ ...,\\
\Gamma^{1}_{20}=& & \frac{1}{r} \left[ c_{,u} ( c_{,\theta} +
2c\cot\theta) - M_{,\theta} \right] \nonumber \\&-&
\frac{1}{2r^{2}} \left[ 2c_{,u}(cc_{,\theta} + 2N) + N +
N_{,\theta\theta} + (N_{,\theta} - N\cot\theta)\cot\theta \right]
+ ... \nonumber \\&+& \frac{l^2}{r^2}{\left(c_{,\theta}+2c\cot\theta
\right)}_{,u} \\&-&\frac{l^2}{r^3}\left[cc_{,u\theta}
+2c\cot\theta(c_{,u} -2) +2c_{,\theta}(c_{,u}-1)+2N_{,u}+M_{,\theta}
\right]+ ..., \nonumber \\
\Gamma^{1}_{21}=& & \frac{1}{r} (c_{,\theta}
+2c\cot\theta) - \frac{1}{r^{2}} (3N + 3cc_{,\theta} -2c^{2}
\cot\theta) + ... \nonumber \\ &-& \frac{l^2}{r^4}\left(cc_{,\theta}
+2c^2\cot\theta \right)+\frac{l^2}{r^5}c\left(cc_{,\theta}+2N\right)+ ...,\\
\Gamma^{1}_{22}=&-& r(1-c_{,u}) +2M -2c_{,\theta}\cot\theta + c (
1 + 2c_{,u} + 2\cot^{2}\theta) -c_{,\theta\theta} + ... \nonumber
\\&-&\frac{l^4}{2r^3}c_{,u} +... ,\\
\Gamma^{1}_{33}/\sin^{2}\theta=&-& r(1+c_{,u}) +2M + c ( 1 +
2c_{,u} - 2\cot^{2}\theta) -c_{,\theta}\cot\theta + ... \nonumber
\\&+& \frac{l^4}{2r^3}c_{,u} +... ,\\
\Gamma^{2}_{00}=& &\frac{1}{r^{2}}{(c_{,\theta}+2c\cot\theta)}_{,u}
\nonumber \\&-& \frac{1}{r^{3}} \left( 2N_{,u} +3cc_{,u\theta} +
M_{,\theta} + 4cc_{,u}\cot\theta + c_{,u}c_{,\theta} \right) + ...
\nonumber \\&-& \frac{4l^4}{3r^6}{(c_{,\theta}+2c\cot\theta)}_{,u}
+ ... , \\
\Gamma^{2}_{01}=& & \frac{N}{r^{4}} + ... + \frac{l^2}{r^5}
{(c_{,\theta}+2c\cot\theta)}_{,u} + ... , \\
\Gamma^{2}_{02}=& & \frac{c_{,u}}{r} -\frac{1}{2r^{3}}(c^{2}c_{,u}
- 2C_{,u}) + ... \nonumber \\&+&\frac{l^2}{r^3}(c_{,u}-1)+
\frac{l^2}{r^4}(2M+c)+ ..., \\
\Gamma^{2}_{10}=& & \frac{N}{r^{4}} + ... - \frac{l^2}{r^5}
{(c_{,\theta}+2c\cot\theta)}_{,u} + ... , \\
\Gamma^{2}_{12}=& & \frac{1}{r} -\frac{c}{r^{2}} -
\frac{1}{2r^{4}}(6C-c^{2}) + ... \nonumber \\&-&\frac{l^2}{r^3}+
\frac{l^2}{r^4}c+ ..., \\
\Gamma^{2}_{20}=& & \frac{c_{,u}}{r} -\frac{1}{2r^{3}}(c^{2}c_{,u}
- 2C_{,u}) + ... \nonumber \\&-&\frac{l^2}{r^3}(c_{,u}-1)-
\frac{l^2}{r^4}(2M+c)+ ..., \\
\Gamma^{2}_{21}=& & \frac{1}{r} -\frac{c}{r^{2}} -
\frac{1}{2r^{4}}(6C-c^{2}) + ... \nonumber \\&+&\frac{l^2}{r^3}-
\frac{l^2}{r^4}c+ ..., \\
\Gamma^{2}_{22}=&-& \frac{2}{r}c\cot\theta + \frac{2}{r^{2}} (N +
cc_{,\theta} + c^{2}\cot\theta) + ... \nonumber \\&-&\frac{l^4}
{6r^5}(c_{,\theta}+2c\cot\theta)+ ...,\\
\Gamma^{2}_{33}=&-& \sin\theta\cos\theta (1 -\frac{2c}{r} -
\frac{2}{r^{2}}N\cot\theta + ... ) \nonumber \\&-&\frac{l^4}
{3r^5}\sin^2\theta(c_{,\theta}+2c\cot\theta)+ ..., \\
\Gamma^{3}_{03}=&-&\frac{c_{,u}}{r} -\frac{C_{,u}}{r^{3}} +
\frac{c^{2}c_{,u}}{2r^{3}} + ... \nonumber \\&-& \frac{l^2}{r^3}
(c_{,u}+1)+ \frac{l^2}{r^4}(2M-c)+ ..., \\
\Gamma^{3}_{13}=& & \frac{1}{r} + \frac{c}{r^{2}}
+\frac{3C}{r^{4}} - \frac{c^{3}}{2r^{4}} + ... \nonumber \\&-&
\frac{l^2}{r^3}-\frac{l^2}{r^4}c+ ..., \\
\Gamma^{3}_{23}=& & \cot\theta -\frac{c_{,\theta}}{r} -
\frac{C_{,\theta}}{r^{3}} + \frac{c^{2}c_{,\theta}}{2r^{3}} + ...
\nonumber \\&+&\frac{l^2}{r^3}(c_{,\theta}+2c\cot\theta)+
\frac{2l^2}{r^4}(N+c^2\cot\theta) + ... ,\\
\Gamma^{3}_{30}=&-&\frac{c_{,u}}{r} -\frac{C_{,u}}{r^{3}} +
\frac{c^{2}c_{,u}}{2r^{3}} + ... \nonumber \\&+& \frac{l^2}{r^3}
(c_{,u}+1)- \frac{l^2}{r^4}(2M-c)+ ..., \\
\Gamma^{3}_{31}=& & \frac{1}{r} + \frac{c}{r^{2}}
+\frac{3C}{r^{4}} - \frac{c^{3}}{2r^{4}} + ... \nonumber \\&+&
\frac{l^2}{r^3}+\frac{l^2}{r^4}c+ ..., \\
\Gamma^{3}_{32}=& & \cot\theta -\frac{c_{,\theta}}{r} -
\frac{C_{,\theta}}{r^{3}} + \frac{c^{2}c_{,\theta}}{2r^{3}} + ...
\nonumber \\&-&\frac{l^2}{r^3}(c_{,\theta}+2c\cot\theta)-
\frac{2l^2}{r^4}(N+c^2\cot\theta) + ... ,
\end{eqnarray}

\end{document}